\documentclass[12pt,titlepage]{article}
\pdfoutput=1

\usepackage[utf8]{inputenc} 
\usepackage{color}
\usepackage{graphicx}
\graphicspath{{figures/}}
\usepackage{tabularx}
\usepackage{threeparttable}
\usepackage{amssymb}
\usepackage{amsmath}
\usepackage{amsthm}
\usepackage{MnSymbol}
\usepackage{authblk}
\usepackage{enumitem}
\usepackage[margin=1in]{geometry}
\usepackage{acronym}
\usepackage{wrapfig}
\usepackage[hidelinks]{hyperref}
\usepackage[sort&compress,numbers]{natbib}
\providecommand{\keywords}[1]{ \small \textbf{\textit{Keywords---}} #1 }

\theoremstyle{definition}
\newtheorem{example}{Example}[section]

\newcommand{\PNNLaffil}{$1$}
\newcommand{\LBTaffil}{$2$}
\newcommand{\PIaffil}{$3$}

\acrodef{AChE}{acetylcholinesterase}
\acrodef{APBS}{Adaptive Poisson-Boltzmann Solver}
\acrodef{DFT}{density functional theory}
\acrodef{GUI}{graphical user interface}
\acrodef{MD}{molecular dynamics}
\acrodef{PB}{Poisson-Boltzmann}
\acrodef{PDB}{Protein DataBank}
\acrodef{Rh-LmrR*}{lactococcal multidrug resistance regulator}
\acrodef{VR}{virtual reality}
\acrodef{AR}{augmented reality}

\begin{document}

\title{Visualizing biomolecular electrostatics in virtual reality with UnityMol-APBS}

\author[\PNNLaffil]{Joseph~Laureanti}
\author[\PNNLaffil]{Juan~Brandi}
\author[\PNNLaffil]{Elvis~Offor}
\author[\PNNLaffil]{David~Engel}
\author[\PNNLaffil]{Robert~Rallo}
\author[\PNNLaffil]{Bojana~Ginovska}
\author[\LBTaffil]{Xavier~Martinez}
\author[\LBTaffil]{Marc~Baaden}
\author[\PIaffil]{Nathan~A.~Baker}

\affil[\PNNLaffil]{Pacific Northwest National Laboratory, Washington, USA.}
\affil[\LBTaffil]{CNRS, Universit\'{e} de Paris, UPR 9080, Laboratoire de Biochimie Th\'{e}orique, 13 rue Pierre et Marie Curie, F-75005, Paris, France.
Institut de Biologie Physico-Chimique-Fondation Edmond de Rothschild, PSL Research University, Paris, France.}
\affil[\PIaffil]{To whom correspondence should be addressed.
Advanced Computing, Mathematics, and Data Division; Pacific Northwest National Laboratory; Richland, WA 99352, USA.
Division of Applied Mathematics; Brown University; Providence, RI 02912, USA. Email:~\href{mailto:nathan.baker@pnnl.gov}{nathan.baker@pnnl.gov}}

\maketitle

%\doublespacing

\begin{abstract}

Virtual reality is a powerful tool with the ability to immerse a user within a completely external environment. 
This immersion is particularly useful when visualizing and analyzing interactions between small organic molecules, molecular inorganic complexes, and biomolecular systems such as redox proteins and enzymes. 
% The biomedical community has direct interest to the interactions between small molecules and protein systems. 
A common tool used in the biomedical community to analyze such interactions is the \ac{APBS} software, which was developed to solve the equations of continuum electrostatics for large biomolecular assemblages. 
Numerous applications exist for using \ac{APBS} in the biomedical community including analysis of protein ligand interactions and \ac{APBS} has enjoyed widespread adoption throughout the biomedical community.
Currently, typical use of the full \ac{APBS} toolset is completed via the command line followed by visualization using a variety of two-dimensional external molecular visualization software. 
This process has inherent limitations:  visualization of three-dimensional objects using a two-dimensional interface masks important information within the depth component. 
Herein, we have developed a single application, UnityMol-\ac{APBS}, that provides a dual experience where users can utilize the full range of the \ac{APBS} toolset, without the use of a command line interface, by use of a simple \ac{GUI} for either a standard desktop or immersive virtual reality experience. 

    \begin{center}
        \includegraphics[width=.45\textwidth]{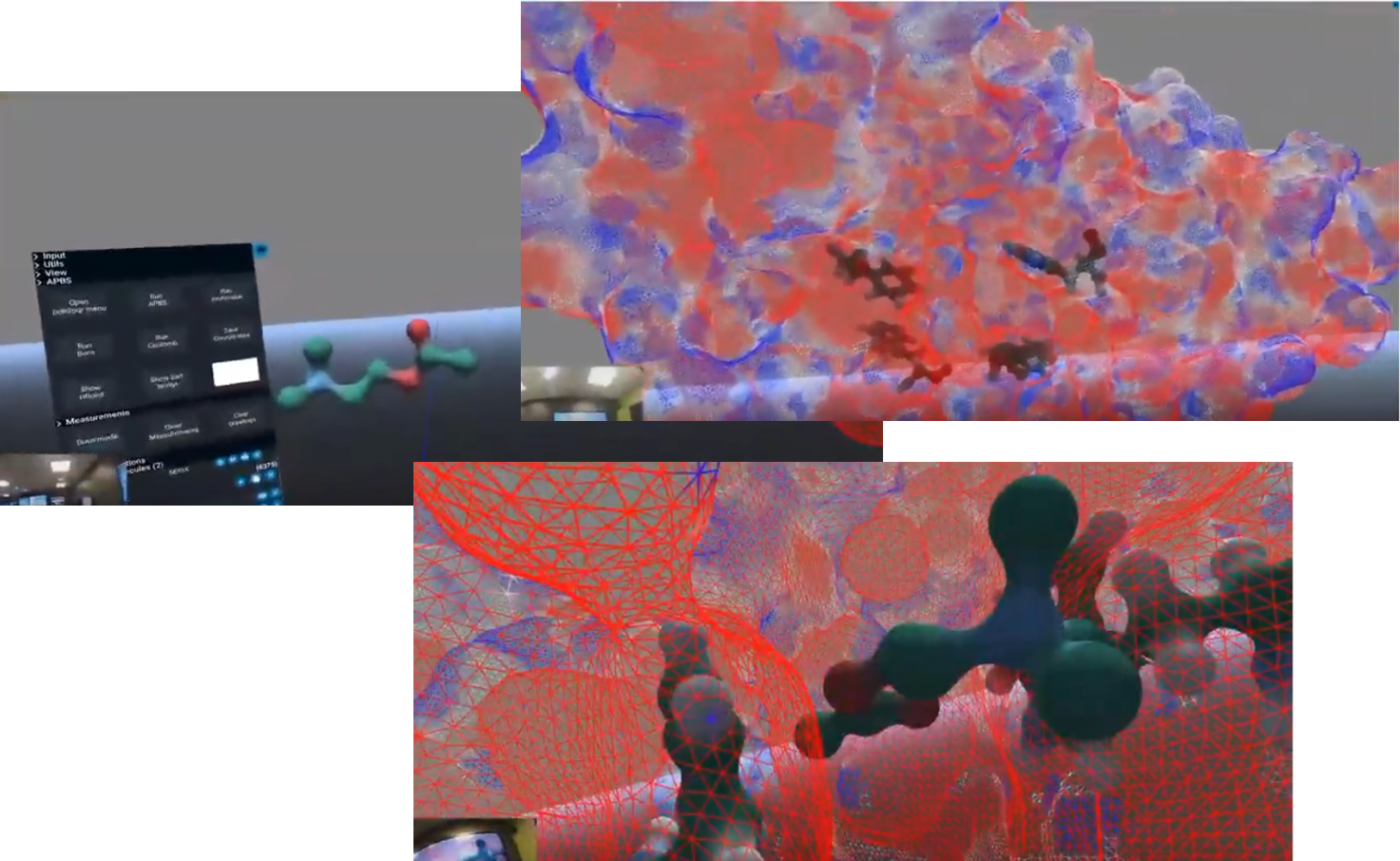}
    \end{center}  
    
\end{abstract}


\section{Introduction} \label{sec:intro}
Understanding and predicting biomolecular processes requires a basic knowledge of the electrostatic interactions possible within a molecular system \cite{Fried2017, Jurrus2018, Dolinsky2007}. 
Many programs\footnote{See \url{https://en.wikipedia.org/wiki/List_of_molecular_graphics_systems} for a partial list of molecular visualization programs.}---such as PyMOL \cite{PyMOL}, VMD \cite{Humphrey1996}, Chimera \cite{Pettersen2004}, and UnityMol \cite{Doutreligne2014}---provide platforms for the visualization and analysis of protein systems.
% Each of these platforms provides varying degrees of methods to calculate and visualize electrostatic information.
However, visualization of molecular systems is currently largely completed using a variety of tools that all contain a common, and inherent limitation: three-dimensional objects are observed using a two-dimensional interface. 
Using a two-dimensional interface to manipulate and understand three-dimensional objects removes an important depth component that is crucial to understanding spatial resolution within a three-dimensional object. 
Furthermore, manipulation of multiple objects in a three-dimensional space is currently performed using a cumbersome combination of mouse or track-pad and keyboard interactions. 
Interacting with a three-dimensional space via a two-dimensional interface is a limitation when attempting to simultaneously investigate multiple objects such as protein-protein, protein-cofactor, or enzyme-substrate interactions.
In this paper, we describe a new UnityMol-\ac{APBS} \ac{VR} platform for visualizing these three-dimensional systems.
We will show brief examples of how a student or researcher would use the \ac{APBS} and PDB2PQR tools to generate and visualize electrostatic information and describe how this use is facilitated by the \ac{VR} interface.
% Finally, we show how the \ac{APBS} toolset can be utilized with custom force field parameters to investigate systems employing non-standard residues.

\subsection{Molecular visualization with UnityMol-APBS} \label{sec:UM}
Many popular platforms exist to visualize molecular systems via two-dimensional environments; e.g., PyMOL, VMD, Chimera, and UnityMol. 
% Each of these platforms fulfills various needs amongst the molecular visualization community. 
Several new visualization methodologies have been implemented to provide an \ac{AR} or \ac{VR} interface \cite{Sommer2018, Nanome, OConner2018, ChimeraX, Educhem, MoleculE}. 
\ac{VR} interfaces---as provided by UnityMol, Chimera (see ChimeraX) \cite{Goddard2018UCSFCM}, Nanome \cite{Nanome}, and Nano Simbox \cite{OConnor2019}---allow students or researchers the ability to be completely immersed within a given molecular system, thereby facilitating insight into interactions between proteins and enzymes, small organic molecules, and synthetic organometallic complexes.
Single-user \ac{VR} software varies from viewing-only applications such as Educhem-VR \cite{Educhem} to tools that provide various degrees of user interaction both in a standard desktop environment and using a \ac{VR} interface; e.g., UnityMol-APBS and ChimeraX \cite{Goddard2018UCSFCM}. 
Collaborative \ac{VR} software includes Nanome \cite{Nanome}, a collaborative \ac{VR} environment for drug discovery, and Nano Simbox \cite{OConnor2019}, a collaborative \ac{VR} environment for interacting with \ac{MD} simulations. 
As noted by the team behind Nano Simbox, \ac{VR} environments have been shown to increase the rate of scientific discovery. 
For example, the developers of Nano Simbox attempted to quantify the positive impact of using a virtual environment for general molecular visualizations and manipulations and found a 10-fold increase in productivity \cite{OConnor2019}.

UnityMol-APBS has been developed using the Unity game engine for both standard desktop (Windows, Mac OS X, and Linux) and \ac{VR} (Windows-only) interfaces.
A Python terminal, implemented using IronPython, has been incorporated to provide an improved user experience through additional command-line functionality.
The original UnityMol software was introduced in 2013 and was developed by Marc Baaden and his research group at CNRS in the Institute of Physico-Chemical Biology.
UnityMol is available free of charge and can be downloaded from a repository at SourceForge\footnote{UnityMol can be downloaded at \url{https://sourceforge.net/projects/unitymol}.}.
The UnityMol-APBS software described in this paper is derived from UnityMol.
% Capabilities of UnityMol-APBS include customization of the visual representation of molecular systems, the ability to concomitantly visualize and manipulate multiple molecular structures, an integrated trajectory player compatible with GROMACS trajectories, and calculation of electrostatic properties using PDB2PQR and \ac{APBS} tools. 
% The virtual world of UnityMol-APBS allows the direct generation of high-quality images to export publication quality figures from within this virtual space.

\subsection{Molecular electrostatics with \ac{APBS}} \label{sec:apbs}
As discussed in our article from the previous {\textit{Tools for Protein Science}} special issue \cite{Jurrus2018}, \ac{APBS} is one of several software packages for calculating electrostatic properties of biomolecular systems\footnote{The \ac{APBS} software can be downloaded or used via the web at \url{http://www.poissonboltzmann.org}.}.
The \ac{APBS} software was developed to solve the \ac{PB} equation \cite{Fixman1979, Grochowski2008, Lamm2003} for the calculation of biomolecular solvation properties and electrostatic interactions:
\begin{equation}
	-\nabla \cdot \epsilon \nabla \phi - \sum_i^M c_i q_i e^{-\beta \left(q_i \phi + V_i \right)} = \rho
	\label{eqn:pbe}
\end{equation}
which is solved on a domain $\Omega \subseteq \mathbb{R}^3$ with $\phi$ specified on the boundary of that domain.
In Eq.~\ref{eqn:pbe}, $\phi: \Omega \mapsto \mathbb{R}$ is the electrostatic potential, $\epsilon: \Omega \mapsto [\varepsilon_u, \varepsilon_v]$ is a dielectric coefficient function that ranges between solute $\varepsilon_u > 0$ and solvent $\varepsilon_v > 0$ dielectric values, and $\rho: \Omega \mapsto \mathbb{R}$ is a charge distribution function.
For each mobile ion species, $i=1,\ldots,M$, $q_i \in \mathbb{R}$ is the charge, $c_i > 0$ is the concentration, and $V_i: \Omega \mapsto \mathbb{R}$ is the steric ion-solute interaction potential.
Finally, $\beta=\left( kT \right)^{-1} > 0$ is the inverse thermal energy where $k$ is the Boltzmann constant and $T$ is the temperature.

The main input to \ac{APBS} is a PQR file that specifies the position in $\Omega$, charge values for $\rho$, and radii (used to construct $\epsilon$ and $V_i$) for each atom. 
The PDB2PQR software was developed to expedite the preparation of input files for analysis by \ac{APBS} \cite{Dolinsky2007, Dolinsky2004}. 
Functions of the PDB2PQR software includes reading a PDB file, converting to the PQR format, repairing missing heavy atoms, optimizing titration states, adding missing hydrogen atoms, assigning radius and charge parameters, and automatically preparing subsequent APBS electrostatic calculations.

\ac{APBS} and PDB2PQR can be used via command-line operations, through popular molecular visualization software (Chimera, PyMOL, VMD), or via a web-based interface.
As described in this article, \ac{APBS} and PDB2PQR tools have now also been incorporated into UnityMol-APBS with three primary goals:
\begin{itemize}
    \item offer an intuitive visual interface to \ac{APBS} functionality without the requirement of compiling,
    \item enable users to employ one application to calculate and immediately visualize and compare multiple results in a completely immersive virtual reality experience, and 
    \item provide the terminal commands, that were generated in UnityMol-APBS, to PDB2PQR and \ac{APBS} as input, in a text file for the purposes of reproducibility of results.
\end{itemize}
Installation of the integrated UnityMol-APBS and \ac{APBS} environment is straightforward: both executables should be downloaded from \url{https://github.com/Electrostatics/VR} and installed on the user system using default values\footnote{UnityMol-APBS assumes a default installation of \ac{APBS} at \texttt{C:$\backslash$APBS$\_$PDB2PQR} on Windows systems.}.
%Installation of the integrated UnityMol-APBS and \ac{APBS} environment is straightforward: both executables should be downloaded and installed on the user system using default values\footnote{UnityMol-APBS assumes a default installation of \ac{APBS} at \texttt{C:$\backslash$APBS} on Windows systems.}.

\section{Methods and results} \label{sec:methods} \label{sec:viz}
\begin{figure}
    \centering
    \includegraphics[width=0.75\linewidth]{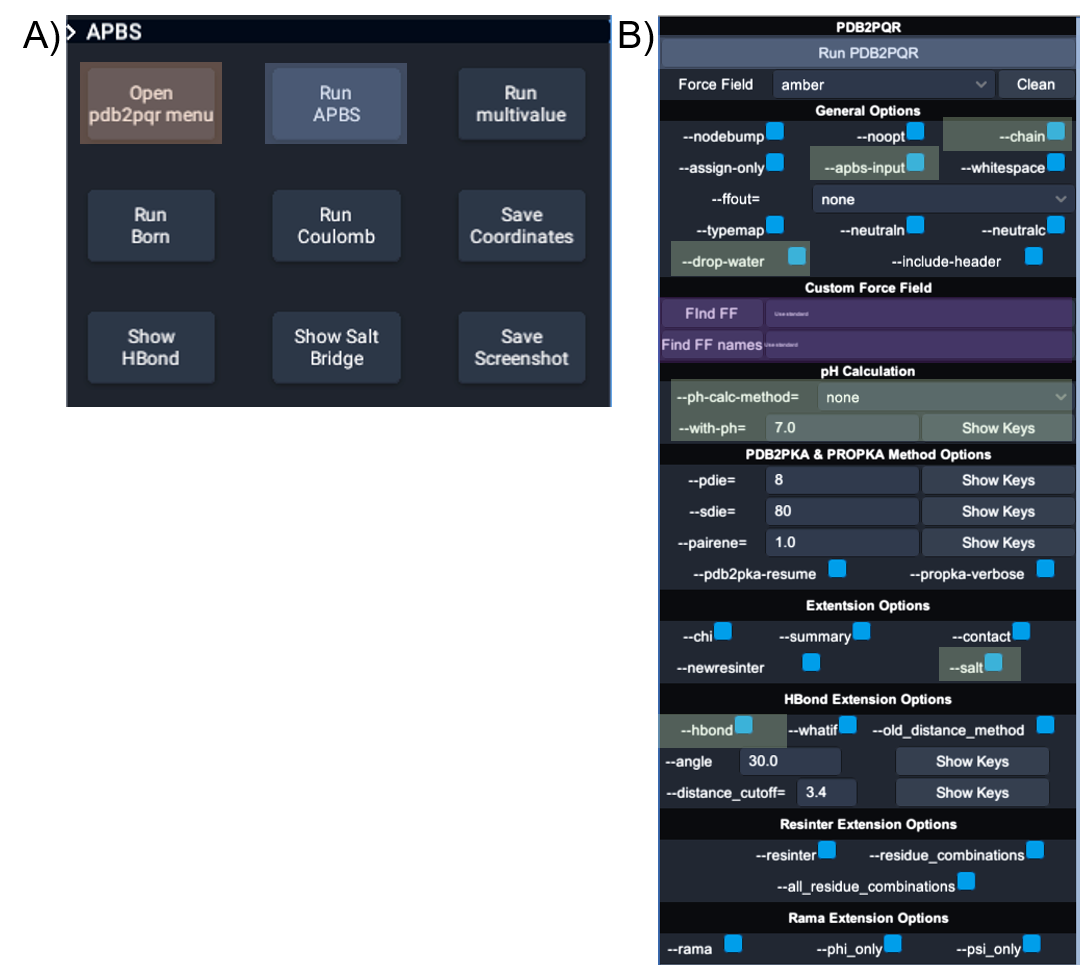}
    \caption{UnityMol-APBS user interface for PDB2PQR and APBS.
    (A)~The main UnityMol-APBS menu; orange box highlights the two buttons used to open the APBS and PDB2PQR tools.
    (B)~The main menu for interactions with APBS and PDB2PQR.
    Blue boxes show the buttons to launch PDB2PQR and APBS executables, green boxes show the location of the options used for producing the images in Figures \ref{fig:ache} and \ref{fig:lmrr-elec}, and the purple boxes highlight the two input fields required to use custom force fields and custom residue names.}
    \label{fig:ui}
\end{figure}

The full \ac{APBS} toolset, including PDB2PQR, was integrated with UnityMol-APBS to allow users access the functionality of this software without the requirement of manually compiling the tools and using terminal commands. 
% This directly enables rapid access of electrostatic information and analysis to a broad range of students. 
UnityMol was specifically chosen for integration with \ac{APBS} due to its high-quality visual representations, ease of interactions with multiple molecular objects, powerful tools for visual manipulations, and the ability to switch between a desktop and a virtual environment. 
All \ac{APBS} tools are directly accessed from within UnityMol-APBS (see Figure~\ref{fig:ui}) and can be used from within the desktop application or a virtual environment to allow immediate comparison of many structural properties including titration states, electrostatic surface potentials, electrostatic field lines, hydrogen bonding interactions, and salt bridge formation.
%%%

The examples below illustrate how to visualize the properties of proteins with natural and unnatural residues.
Values and methodologies used in the examples are intended as a demonstration and are not necessarily the recommended values for all applications.
Please refer to the PDB2PQR and \ac{APBS} documentation when using APBS for your own applications. 
The HTC Vive \ac{VR} system was used for the examples shown in this paper; however, several \ac{VR} systems are supported including: Oculus CV1, DK2, and Rift S.
Both the desktop and \ac{VR} environments of UnityMol-APBS save all user commands as a Python script that can be loaded later to reproduce or reuse the visualization workflow.
This greatly reduces the time required to begin working from where a researcher left off, or when a student is loading a scene that an instructor has previously prepared.

\subsection{Analyzing \ac{AChE} electrostatic surface potentials and field lines} \label{sec:ex-ache}
\begin{figure} 
	\begin{center}
		\includegraphics[width=.95\linewidth]{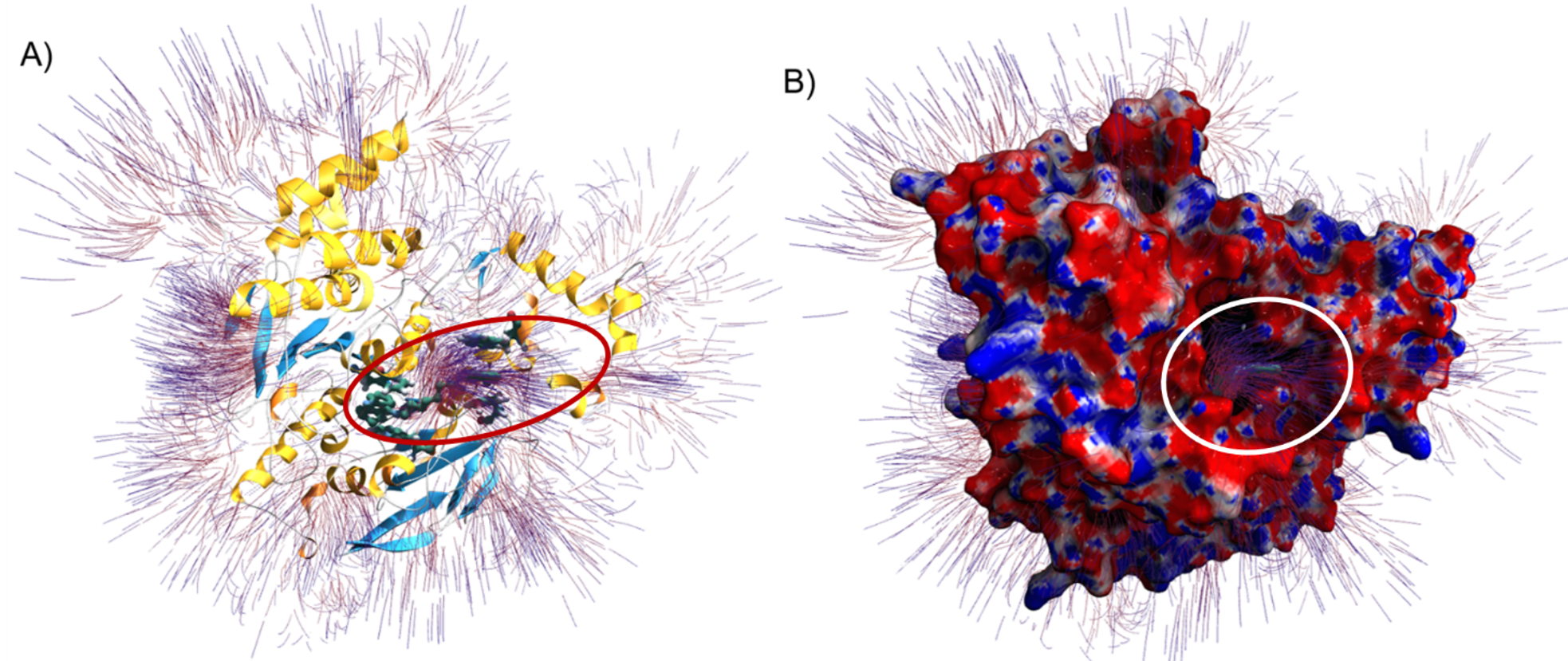}
		\caption{Electrostatic surface potential and field lines of \textit{Torpedo californica} \ac{AChE} (\ac{PDB} ID 5EI5) with bound alkylene-linked bis-tacrine.
        (A)~Electrostatic field lines and protein secondary structure shown with alpha helices (yellow), beta sheets (blue), and random coils (white). Residues Tyr70, Trp84, Trp279, and Phe330 are shown interacting with alkylene-linked bis-tacrine via hydrogen bonding and $\pi$-$\pi$ stacking interactions.
        The red oval highlights the potential gradient.
        (B)~AChE surface model with field lines and mapped electrostatic surface potentials shown with neutral, negative, and positive charges in white, red, and blue, respectively.
        Field lines are calculated from a gradient (value of 0.2) and depicted with the starting points in red and the ending points in blue.
        The orientation is the same in Figures A and B, where the alkylene-linked bis-tacrine can be seen occupying the catalytic gorge.
        The white circle highlights the potential gradient exiting the catalytic gorge.} 
        \label{fig:ache}
	\end{center}  
\end{figure}

\textit{Torpedo californica} \ac{AChE} is an enzyme that operates incredibly fast with high selectivity, in part, due to its electrostatic properties \cite{Quinn1987}.
Catalysis occurs at an active site buried within the enzyme.
Electric fields help guide positively charged substrates from the enzyme exterior through a narrow channel to the buried active site as shown in the UnityMol-APBS visualization in Figure \ref{fig:ache}. 
As such, this enzyme is a good example for exploring electrostatic surface potentials and electrostatic field lines in an immersive and virtual experience. 
The instructions provided here for the \ac{AChE} analysis can be used as a template for other uses of UnityMol-APBS.
The Appendix includes step-by-step instructions for this example (Example~\ref{ex:ache-steps}) as well as the output Python code (Example~\ref{ex:ache-python}), which can be used by a user to instantly load our results.

Structural information for \textit{Torpedo californica} \ac{AChE} was obtained from \ac{PDB} entry 5EI5, via the fetch command in UnityMol-APBS. 
Once loaded, the default force field parameters for AMBER \cite{Case2005} were chosen and amino acid protonation states were estimated using PROPKA at pH 7.0. 
PROPKA is a heuristic method for computing p$K_a$ values by incorporating effects due to desolvation, hydrogen bonding, and charge–charge interactions \cite{Sondergaard2011}. 
Figure \ref{fig:ui} highlights the options used with PDB2PQR: --chain, --apbs-input, --summary, --drop-water, --salt, and --hbond, to retain the chain information, prepare an input file for APBS, print a summary, remove water molecules, write salt bridge interactions, and write hydrogen bonding interactions, respectively. 
UnityMol-APBS creates and saves all files in the destination directory \texttt{C:\textbackslash APBS\_PDB2PQR\textbackslash OutputFiles}. Upon running the PDB2PQR tools, the newly created .PQR file is immediately loaded for the user. 
Hydrogen bonding and salt bridge information are written as \texttt{.hbonds} and \texttt{.salt} formatted files; this information can be read and visualized directly from within UnityMol-APBS as independent selections. 
This allows users to color the interactions as desired. 

Default parameters for \ac{APBS} Poisson-Boltzmann calculations were selected through the Unity\-Mol-APBS interface. 
\ac{APBS} writes electrostatic potential output in OpenDX format that is directly loaded by UnityMol-APBS after completing the calculation. 
The potential gradient within the OpenDX file allows UnityMol-APBS to color the molecular surface based on electrostatic potential and show the associated electrostatic field lines, Figure \ref{fig:ache}. 
Highlighted in Figure 2 is the strong electric field originating in the \ac{AChE} active site channel. 
Due to the confinement of the catalytic gorge, this channel is extremely challenging to explore with traditional molecular visualization methods. 
Additionally, manually moving a ligand through the catalytic gorge to analyze the residues that could interact is not a task that can be completed easily in 2D visualization environments.
Completing this task in \ac{VR} is much easier as the depth component is not obscured due to the inherent limitations associated with viewing a three-dimensional space on a two-dimensional screen. 
Furthermore, the virtual environment more easily allows the user to simultaneously manipulate both the enzyme and the substrate to provide optimized viewing angles. 
Animated electrostatic field lines in UnityMol-APBS represent the electrostatic potential gradient and the following properties can be manipulated: movement speed of the lines, width of lines, length of the line, and the gradient used to calculate the lines.
%An example of the electrostatic surface potential and electrostatic field lines of \textit{Torpedo californica} \ac{AChE} is shown in the supplementary information video available at \url{https://www.youtube.com/watch?v=mnMDt_Z9pf0}.
An example of the electrostatic surface potential and electrostatic field lines of \textit{Torpedo californica} \ac{AChE} is shown in the supplementary information video, which can be accessed at \url{https://github.com/Electrostatics/VR}.
Also, included in the video is an example of a user manipulating two objects (substrate and \ac{AChE}) simultaneously in a virtual space manually traversing the substrate through the catalytic gorge of \ac{AChE}.

\subsection{Using custom force fields to analyze an artificial enzyme: Rho\-di\-um-containing Lactococcal Multidrug Resistance Regulator (Rh-LmrR*)} \label{sec:ex-lmrr}
\begin{figure}
	\begin{center}
		\includegraphics[width=.95\linewidth]{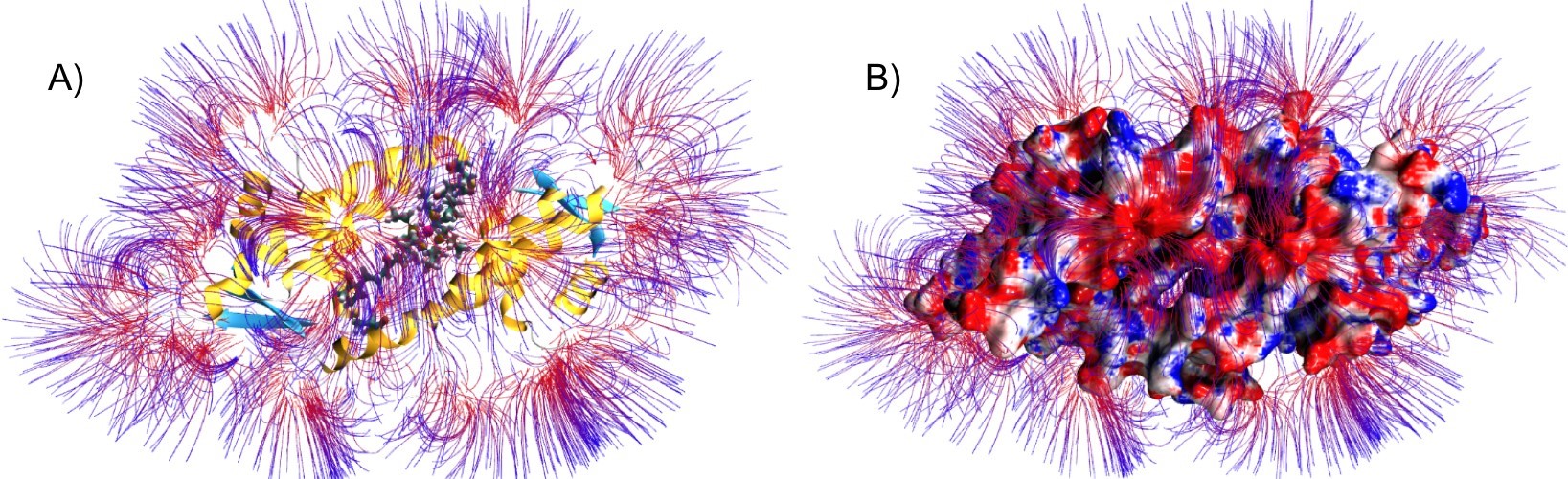}
        \caption{Single conformation from a \ac{MD} simulation of \ac{Rh-LmrR*} showing (A) the protein secondary structure and the immobilized Rh-complex. (B) electrostatic surface potential and field lines. Color scheme is identical to Figure \ref{fig:ache}. }
        \label{fig:lmrr-elec}
     \end{center}  
\end{figure}

The electrostatic field lines an electrostatic surface potentials of an artificial metalloenzyme capable of hydrogenating CO$_2$ to a liquid fuel, formate, is shown in Figure \ref{fig:lmrr-elec}~\cite{Laureanti2019}. 
This system, Rh-LmrR (where LmrR is the Lactococcal multidrug resistance regulator and Rh refers to an organometallic rhodium-(bis)diphosphine complex), was chosen as the second example to illustrate the use of custom force fields with UnityMol-APBS. 
The artificial metalloenzyme contains a covalently anchored Rh(bis)diphosphine cofactor that cannot be analyzed using standard force field parameters due to the non-standard rhodium and phosphorus atoms, as well as the associated ligand framework. 
In this example, we show how to input externally generated force field data, analyze hydrogen bonding interactions, identify salt bridge locations, calculate the electrostatic potential at multiple atoms using the APBS \text{multivalue} tool, as well as visualize the electrostatic surface potential and electrostatic field lines.

Our methods to prepare the custom force fields, using tools outside of UnityMol-APBS, are briefly described here. 
Structural data for the artificial metalloenzyme was first obtained from crystallographic data (PDB ID 6DO0) with further minimization via density functional theory (DFT) calculations and conformational sampling using MD simulations in GROMACS~\cite{Kumari2014}. 
The resulting PDB file for Rh-LmrR was loaded into UnityMol-APBS and PDB2PQR was used to prepare the structure for \ac{APBS} calculations as described in the AChE example above. 
The parameters for the Rh-(bis)diphosphine complex do not exist in any of the included standard libraries, therefore custom force field parameters for Rh-LmrR were explicitly input to UnityMol-APBS.
Calculations for DFT were performed in NWChem 6.5~\cite{NWChem} using the B3LYP functional~\cite{B3LYPa, B3LYPb} and 6-31G* basis set~\cite{631Ga, 631Gb} for all atoms except for Rh, which used a large-core correlation-consistent pseudopotential basis set~\cite{Martin2001}. 
The geometries were optimized in the gas phase and verified as minima using Hessian calculation in a rigid-rotor, harmonic approximation.
For the charge calculation, the complex was capped with methyl groups, constrained to an overall neutral charge, and all equivalent atoms were constrained to equal values.
Charges for atoms other than Rh and P were calculated using the RESP procedure~\cite{RESP}. 
The atomic point charges of the Rh center and phosphorus atoms were obtained using the Bl\"{o}chl scheme~\cite{Blochl1995}, since the RESP scheme resulted in unphysical values. 
Lennard-Jones parameters for each atom were taken from the GAFF force field~\cite{GAFF}, except for Rh which used UFF~\cite{UFF}.

UnityMol-APBS allows custom force fields and residue names to be used by duplicating the base force field files and amending new custom force field and residue name data as described in the APBS-PDB2PQR documentation. 
Once these files are prepared, the data can be loaded via the PDB2PQR menu in UnityMol-APBS. 
The location of the custom force field is provided by use of the {``Find FF''} button, and the custom residue names are found by use of the {``Find FF names''} button, shown highlighted in purple in Figure \ref{fig:ui}.
When the PDB2PQR executable is launched via UnityMol-APBS (highlighted blue in Figure \ref{fig:ui}), the paths to the custom files will be automatically loaded, overriding the force field information chosen at the top of the \ac{GUI}. 
After the PDB2PQR process finishes, the newly generated \texttt{.pqr} file is automatically loaded into UnityMol-APBS as a new selection. 
The APBS executable (highlighted in blue, Figure \ref{fig:ui}) can then be launched to solve the Poisson-Boltzmann equations for continuum electrostatics. 
The newly generated OpenDX file will be automatically loaded into UnityMol-APBS, thus facilitating visualization of the electrostatic surface potential and electrostatic field lines. 

As an example of how to evaluate specific atomic potential values from within UnityMol-APBS, we used the immersive virtual reality interface to compare the electrostatic potential of the four phosphorus atoms and the Rh metal center of the organometallic complex alone in solution to the Rh complex immobilized within the protein scaffold. 
In the original publication~\cite{Laureanti2019}, it was found that the Rh-complex alone in solution was not a competent catalyst for CO$_2$ hydrogenation.
However, catalytic activity was observed once the Rh complex was immobilized in the protein scaffold. 
The protein scaffold imposes conformational restrictions and as this example implies, also perturbs the electrostatic potential around the Rh-P environment. 
This shift in electrostatic potential was observed using the \ac{APBS} \texttt{multivalue} tool via UnityMol-APBS. 
To complete this task without imposing influences from the grid, we used the following workflow
% (see Table \ref{tab:workflow} for the values obtained at each step)
to accurately calculate potentials while removing ``self-energy'' artifacts from the grid.
First, two initial \texttt{.dx} files containing the calculated electrostatic potential map were created from the respective (\ac{APBS}) input files for the Rh-complex and the artificial enzyme using a protein dielectric (\texttt{pdie}) of 2 and a solvent dielectric (\texttt{sdie}) of 78.4. 
Second, an additional \texttt{.dx} file was created for each system with a homogeneous dielectric of 2 (\texttt{pdie=sdie=2}). 
Third, the data from each \texttt{.dx} file was then used to generate potential values from the \texttt{multivalue} tool by selecting the desired atoms (one Rh atom and four P atoms) of the active site. 
Fourth, the reaction field potential values were calculated by subtracting the homogeneous-dielectric values (second step) from the inhomogeneous-dielectric values (first step) to remove self-energy grid artifacts in the calculation.
Fifth, the APBS \texttt{coulomb} tool was used to analytically calculate the electrostatic potential at each atom, assuming a homogeneous environment of dielectric 2. 
The analytically calculated Coulomb potential results were added to the reaction field potential results to give the final potential value at each atom site.
From these results, we observed that all four phosphorus atoms were perturbed upon immobilization with an average shift of $165 \pm 15$ mV relative to the Rh-complex alone in solution. 
The Rh center shifted -170 mV upon immobilization within the protein environment. 
% These results are merely an example of how to use the multivalue tool. 
% As each system investigated, Rh-complex alone and Rh-LmrR are examples from single snapshots in time, they are not representative of these systems under catalytic conditions and are not representative of an actual explanation for the cause of introducing catalytic activity in the Rh-LmrR system.

% \begin{table}[h!]
%     \centering
%     \caption{Values obtained for the artificial enzyme from the workflow for removing grid influences using the Multivalue and Coloumb tools.}
%     \label{tab:workflow}
%     \begin{tabular}{ | c | c | c | c | c | c | }
%     \hline
%     & Multivalue &  & Multivalue &  &    \\
%     & output 1 &    & output 2 &    &    \\ 
% 	\hline
%     & pdie=2, & pdie=2, & multivalue output 1 - &   & Sum column D \\
%     Atom & sdie=78.4 & sdie=2	& multivalue output 2 &	Coulomb / 2	& and column E \\
%     \hline
%     P	 & -6.22E-01 & -1.44E+02 & 1.43E+02 & -2.41E+01 & 1.19E+02 \\
%     \hline
%     P1 & 3.20E+01  & -8.70E+01 & 1.19E+02 & -3.39E+01	& 8.51E+01 \\ 
%     \hline
%     P2 & 4.44E+01  & -7.29E+01 & 1.17E+02 & -3.47E+01	& 8.26E+01 \\ 
%     \hline
%     P3 & -1.66E+01 & -1.41E+02 & 1.25E+02 & -2.82E+01	& 9.66E+01 \\
%     \hline
%     Rh & 2.88E+02  & 1.58E+02  & 1.30E+02 & 1.40E+01	& 1.45E+02 \\
%     \hline
%     \end{tabular}
% \end{table}

\begin{figure} 
    \centering
    \includegraphics[width=.95\linewidth]{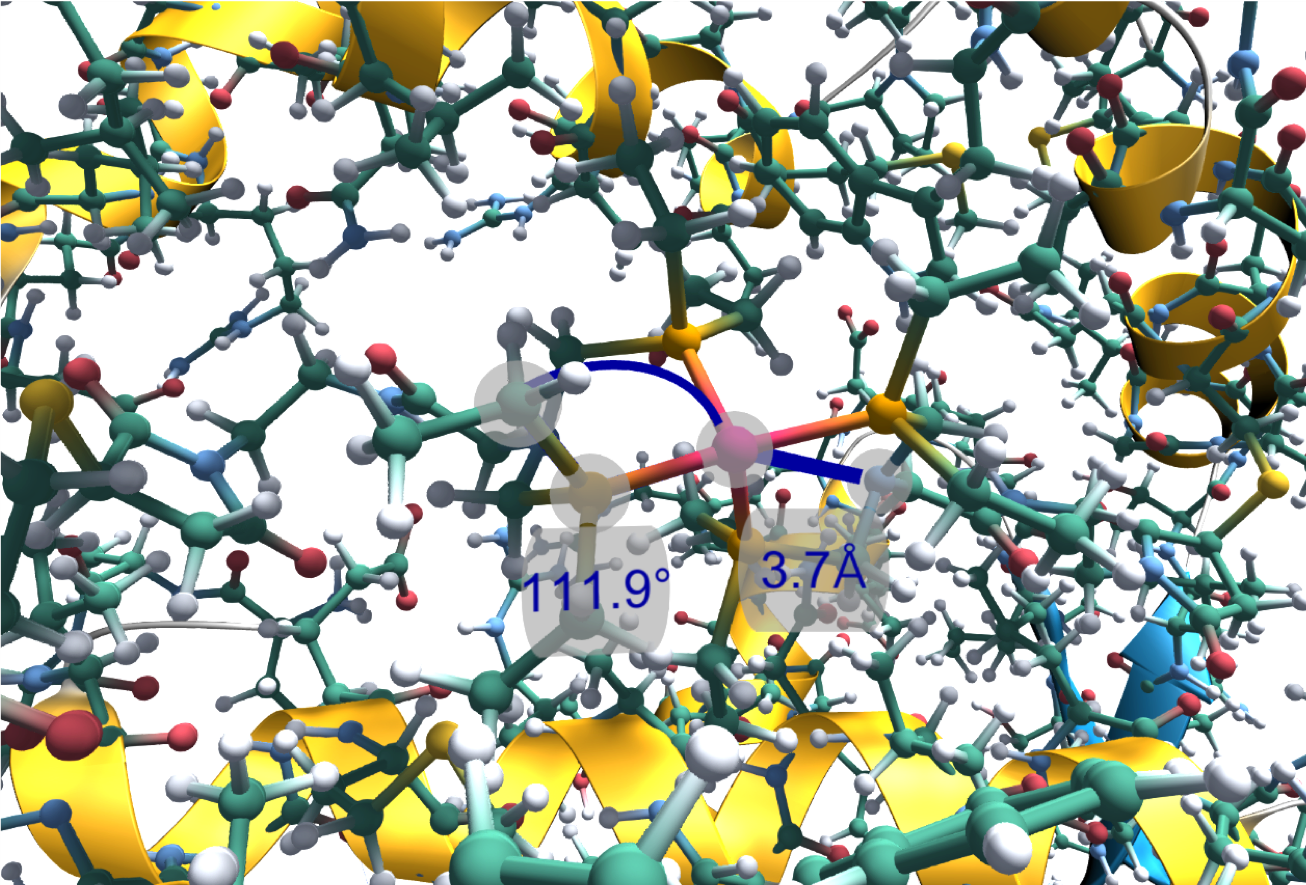}
    \caption{Example of using UnityMol-APBS to explore atomic distances and bond angles.
    Using a virtual environment to complete this task is simpler than a two-dimensional interface.
    Dihedral angles can also be shown directly using the UnityMol-APBS interface.
    Distance measurement line and angle measurement line are shown as blue lines.
    Grey spheres are used to highlight the selected atoms.}
    \label{fig:lmrr-hbond}
\end{figure}

Using \ac{VR} to explore the active site of an artificial enzyme in an immersive environment is extremely helpful when trying to understand the electrostatic environment, spatial constraints, local flexibility, and potentially beneficial or inhibiting interactions of neighboring residues. 
A full immersion experience allows the user to observe the protein surface from within the protein scaffold, a feature that is drastically hindered when using a standard two-dimensional interface. 
As illustrated in Figure \ref{fig:lmrr-hbond}, the \ac{VR} environment is particularly well-suited for measuring distances, angles, and electrostatics potentials at specific sites in the protein (e.g., the Rh metal center). 
The \ac{APBS} multivalue tool can be used to identify the electrostatic potential at specific atoms selected in UnityMol-APBS and is described in further detail in the Appendix (Example~\ref{ex:lmrr-steps}).

\section{Conclusions}
In this newly coupled version of UnityMol-APBS, we have provided a seamless desktop/virtual interface to provide a robust method of interaction with molecular systems. 
The virtual world allows a user to easily manipulate and compare multiple three-dimensional objects in real time, a task that is very much hindered using only a standard desktop interaction. 
Additionally, the virtual interface allows users to quickly prepare input files for the PDB2PQR and APBS toolset by simply reaching out and grabbing desired atoms, which are then automatically prepared by the UnityMol-APBS software and delivered to the APBS tools via command line interactions. 
This dramatically reduces the time required, since the user no longer has to sift through multiple text files to prepare custom input files for the PDB2PQR and APBS toolsets. 
Furthermore, the inherent need to interact with the PDB2PQR and APBS tools through direct command line interactions has been removed. 
All command line inputs are setup using a helpful \ac{GUI} and delivered directly from UnityMol-APBS. 
This greatly increases the availability of the PDB2PQR and APBS toolsets to undergraduate students as well as anyone entering the field of molecular visualization, as users are not required to compile the software before use, nor is there an inherent need to understand command line interactions. 

\begin{figure}
	\begin{center}
		\includegraphics[width=.55\linewidth]{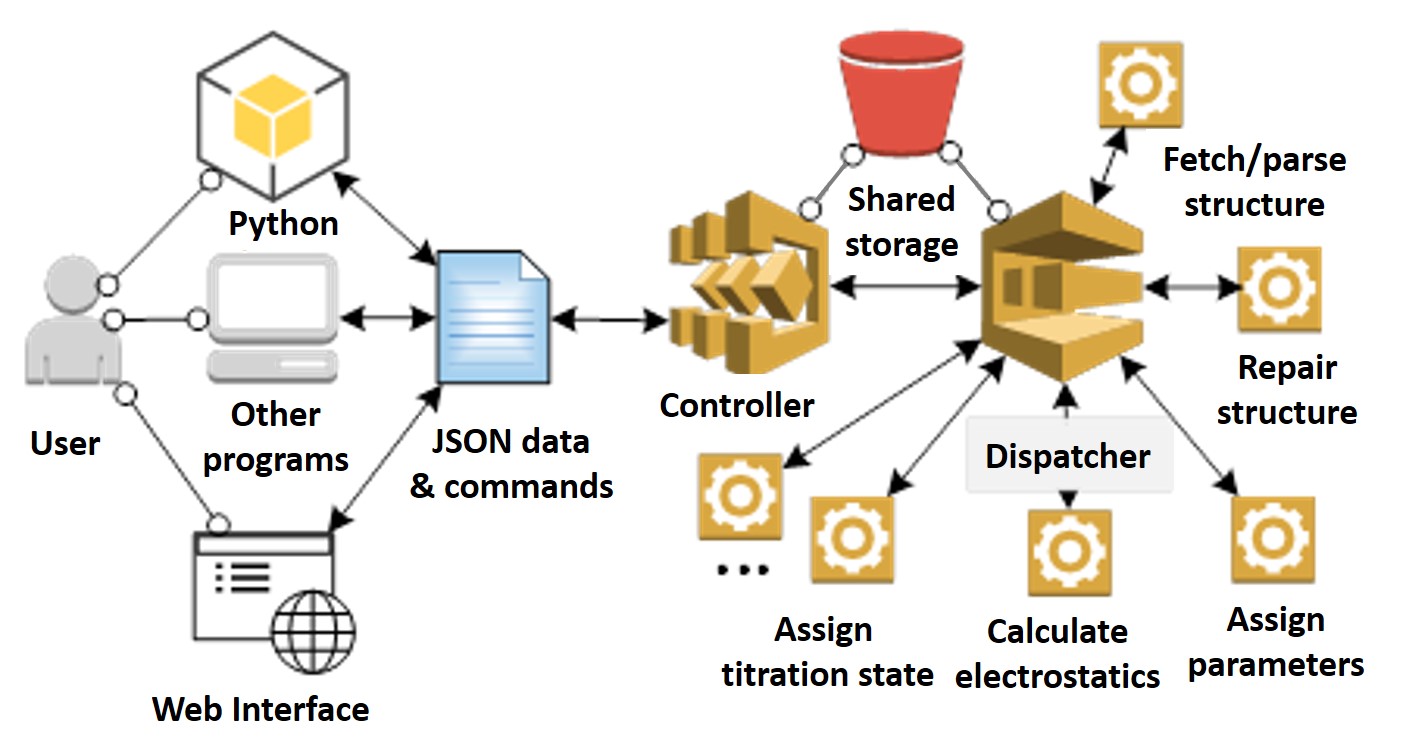}
        \caption{Diagram for redesigned and refactored \ac{APBS} software to enable modular growth.} \label{fig:SWdev}
	\end{center}  
\end{figure}

\ac{APBS} has been in public use for nearly 20 years, thanks to ongoing support by the National Institutes of Health to maintain and update the codes, work with the user base to ensure access and execution of the codes, and improve performance and incorporate new features based on new algorithms and user feedback.
Any software package still in use over that span of time requires continuous updating and refinement to leverage emerging software technologies and improvements in measurement technology that affect the input data.
Therefore, \ac{APBS} has been redesigned, as illustrated in Figure \ref{fig:SWdev}, to ensure the continued availability of this free and scalable software package for biomolecular electrostatics, solvation, and structure assessment/preparation.
The goal of the ongoing software re-engineering effort is to unify the software development into a single code-base that can be easily deployed on desktops, servers and cloud infrastructures.
To achieve this goal, the software has been refactored as a microservices architecture to improve its extensibility and scalability. 
% The PDB2PQR-APBS workflow has been implemented as a collection of loosely coupled components (i.e., microservices) offering lightweight REST interfaces, as illustrated in Figure \ref{fig:Micro}. 
% Microservices are deployed within Linux-based Docker containers to facilitate software maintenance and to ensure application scalability. 
% Kubernetes, which is an open-source framework for container orchestration, is used for application deployment and management. The web-based user interface has also been redesigned to leverage current web development technologies and to provide improved user interaction.  
% This redesign supports continued maintenance of the code as well as facilitate the addition of new features to enhance user experience, incorporate new models and methods, and couple \ac{APBS} with other software such as future versions of UnityMol-APBS and other \ac{VR} visualization software.
The redeveloped software is currently under testing.
Access to this new version of \ac{APBS} will be available through \url{http://www.poissonboltzmann.org}; the underlying developmental code is available at \url{https://github.com/Electrostatics/apbs-rest}.
% %
% \begin{figure}
% 	\begin{center}
% 		\includegraphics[width=.85\linewidth]{APBS-Micro.png}
%         \caption{Design of the new \ac{APBS} web server into a microservice framework.} \label{fig:Micro}
% 	\end{center}  
% \end{figure}

\section*{Acknowledgments}
The authors gratefully acknowledge NIH grant GM069702 for support of \ac{APBS} and PDB2PQR. 
The development of the \ac{VR} methods was supported by the US Department of Energy (DOE), Office of Science, Office of Basic Energy Sciences (BES), Division of Chemical Sciences, Geosciences, \& Biosciences.
MB acknowledges support by the ``Initiative d'Excellence'' program from the French State (Grant ``DYNAMO'', ANR-11-LABX-0011, and ``CACSICE'', ANR-11-EQPX-0008).
PNNL is operated by Battelle for the U.S.\ DOE under contract DE-AC05-76RL01830.

\pagebreak
\appendix

\section{Appendix}

The following examples provide step-by-step directions for the systems described in the main text. 
The source files are also available for download as Supporting Materials for this article at \url{https://github.com/Electrostatics/VR}.

\begin{example}[Step-by-step directions for \ac{AChE} \ac{VR} visualization in Section \ref{sec:ex-ache}]
    \label{ex:ache-steps}
    \mbox{} \
    \begin{enumerate}
        \item Open UnityMol-APBS (\ac{VR} or desktop)
        \item Load \texttt{5ei5.pdb} file
        \item Open PDB2PQR panel
        \item Choose options (examples below) or run the default (default force field is AMBER)
        \begin{description}
            \item[apbs-input] generates input file necessary for \ac{APBS}
            \item[drop-water] removes explicit water molecules from structure
            \item[summary] writes atom names and sequence to a new file
            \item[salt] writes salt bridge donor and acceptor atoms to a new file
            \item[hbond] writes hydrogen bonding donors and acceptors to a new file
            \item The resulting .hbond and .salt files can be loaded as a new selection in UnityMol-APBS
        \end{description}
        \item Select “all(5EI5)” and run PDB2PQR
        \item 5ei5X.pqr is written to a file and is immediately loaded for the user.
        \item Select “all(5EI5)” and run \ac{APBS}
        \item 5ei5X.dx is created and loaded into the selection “all(5EI5X)” automatically
        \item Select the “+” button on the “all(5EI5X)” selection tab, then select “surface”
        \item Select ``color by charge''
        \item Select the ``+'' button on the “all(5EI5X)” selection tab, then select “field lines”
    \end{enumerate}
\end{example}

\pagebreak

\begin{example}[Output Python code from UnityMol-APBS for \ac{AChE} \ac{VR} visualization in Section \ref{sec:ex-ache}]
    \label{ex:ache-python}
    \mbox{} \
    \begin{verbatim}
        load("D:/5EI5_May/5ei5.pdb", True, False)
        load("D:/5EI5_May/5ei5X.pqr", True, False)
        loadDXmap("5ei5X", "D:/5EI5_May/5ei5X.dx")
        showSelection("all(5ei5X)", "fl")
        showSelection("all(5ei5X)", "s")
        colorByCharge("all(5ei5X)", "s") 
        #UnityMol Specific functions ------------ 
        #Save parent position 
        setMolParentTransform(Vector3(169.7430, -26.8734, -168.5344),
                              Vector3(0.0178, 0.0178, 0.0178),
                              Vector3(0.0000, 0.0000, 0.0000),
                              Vector3(0.5442, 1.4145, 1.0517))
    \end{verbatim} 
\end{example}

\pagebreak

\begin{example}[Step-by-step directions for \ac{Rh-LmrR*} \ac{VR} visualization in Section \ref{sec:ex-lmrr}]
    \label{ex:lmrr-steps}
    \mbox{} \
    \begin{enumerate}
        \item Open UnityMol-APBS (\ac{VR} or desktop)
        \item Load \texttt{RhLmrR.pdb} file 
        \item Open PDB2PQR panel
        \item Choose options (examples below) or run the default (default force field is AMBER)
        \begin{description}
            \item[apbs-input] generate input file necessary for \ac{APBS} 
            \item[drop-water] removes explicit water molecules from structure 
            \item[summary] writes atom names and sequence to a new file
            \item[salt] writes salt bridge donor and acceptor atoms to a new file
            \item[hbond] writes hydrogen bonding donors and acceptors to a new file
            \item The resulting .hbond and .salt files can be loaded as a new selection in UnityMol-APBS
        \end{description} 
        \item Click ``Find FF'' (see Figure~\ref{fig:ui}, purple box) and navigate to custom force field file (\texttt{CustomFF.dat})
        \item Click ``Find FF names'' (see Figure~\ref{fig:ui}, purple box) and navigate to custom names file (\texttt{CustomFF\_resNames.names})
        \item Select “all(RhLmrR)” and run PDB2PQR, RhLmrRX.pqr is written to a new file and is loaded automatically
        \item Select “all(RhLmrRX)” and run \ac{APBS}, RhLmrRX.dx is written to a new file and is loaded automatically into the selection “all(RhLmrRX)”
        \item Select the “+” button on the “all(RhLmrRX)” selection tab and select “surface”
        \item Select “color by charge” under the surface selection tab
        \item Select the “+” button on the “all(RhLmrRX)” selection tab and select “field lines”
        \item Select the “New Selection” button under the “Utils” tab
        \item Click “Selection mode” twice to change to “Atom”
        \item Select the four phosphorus atoms and the Rh atom, the new selection should contain five atoms
        \item With the new selection highlighted, click the “multivalue” tool button, under the \ac{APBS} tools tab, to evaluate the electrostatic potential for these five atoms. 
        \begin{description}
            \item Two important text files are created: 1) containing the original data for these atoms from the .pdb file and 2) the output from multivalue. The order of the lines in file 1 are equivalent to that of file 2. This is helpful when keeping track of multiple atoms evaluated with multivalue.
        \end{description} 
    \end{enumerate}
\end{example}

\bibliography{apbs}
% \printbibliography

\end{document}